\documentstyle[12pt,a4wide]{article}

\newcommand\MeV{{\,\rm MeV\/}}
\newcommand\GeV{{\,\rm GeV\/}}
\newcommand\as{{\alpha_s}}
\newcommand{\slD}{\kern1pt/\kern-8pt D}
\newcommand{\slk}{/\kern-6pt k}
\newcommand{\sll}{/\kern-4pt l}
\newcommand{\slp}{p\kern-5pt/}
\newcommand{\slv}{v\kern-5pt\raise1pt\hbox{$\scriptstyle/$}\kern1pt}

\begin{document}
\begin{flushright}
MZ-TH/97-34\\
hep-ph/9710365\\
October 1997
\end{flushright}
\vspace{1truecm}
\begin{center}
{\Large\bf Analyzing QCD Sum Rules for Heavy Baryons}\\[.3truecm]
{\Large\bf at Next-to-Leading Order in {\boldmath$\alpha_S$}
\footnote{Talk given at the IVth International Workshop on Progress in 
  Heavy Quark Physics,\\
  \hbox{\qquad\qquad} Rostock, September 20--22, 1997}}\\[.5truecm]
{\large Stefan Groote}\\[.5truecm]
Institut f\"ur Physik, Johannes-Gutenberg-Universit\"at,\\[.3truecm]
Staudinger Weg 7, D-55099 Mainz, Germany\\[1truecm]
\end{center}
\tableofcontents
\begin{abstract}
In this talk I consider QCD sum rules for the ground state heavy baryons to 
leading order in $1/m_Q$ and at next-to-leading order in $\alpha_S$ within 
the context of Heavy Quark Symmetry. The analysis is done at a fixed scale 
$\mu=1\GeV$. The evolution behaviour of the residues is determined by the 
two-loop anomalous dimension of the heavy baryon currents calculated earlier. 
I compare the diagonal, non-diagonal and constituent type QCD sum rules and 
show that the diagonal sum rules are the most reliable one. As central values 
for the bound state energies I find $m(\Lambda_Q)-m_Q\simeq 760\MeV$ and 
$m(\Sigma_Q)-m_Q\simeq 940\MeV$. The central values for the residues are 
given by $F(\Lambda_Q)\simeq 0.030\GeV^3$ and $F(\Sigma_Q)\simeq 0.038\GeV^3$.
\end{abstract}

\newpage

\section{Introduction}
The correlator of two baryonic currents is the fundamental building block
for stating QCD sum rules for heavy baryons, i.e.\ baryons which contain 
one heavy quark. In this talk I present the calculation of first order 
radiative corrections to QCD sum rules within the Heavy Quark Effective 
Theory (HQET) in the limit of infinite mass for this quark~\cite{rostock1}. 
The calculation of first order radiative corrections~\cite{rostock2,rostock3} 
make sense because of the knowledge of the two-loop anomalous 
dimension~\cite{rostock4}. The interpolating currents for the heavy baryons 
can be written as
\begin{equation}
J_B=[(q_1)^{iT}C\Gamma\tau(q_2)^j]\Gamma'Q^k\varepsilon_{ijk},
\end{equation}
where the index $T$ means transposition, $C$ is the charge conjugation
matrix, $\tau$ is a matrix in flavour space, and $i$, $j$ and $k$ are
color indices. $\Gamma$ and $\Gamma'$ are the light-side and heavy-side
Dirac structures of the current vertices. For each ground-state baryon 
there are two independent current representations~\cite{rostock5} carrying 
the same quantum numbers, for the $\Lambda_Q$-type baryons they are e.g.\ 
given by
\begin{equation}
J_{\Lambda_1}=[(q_1)^{iT}C\tau\gamma_5(q_2)^j]Q^k\varepsilon_{ijk}\quad
  \mbox{and}\quad
J_{\Lambda_2}=[(q_1)^{iT}C\tau\gamma_5\slv(q_2)^j]Q^k\varepsilon_{ijk}.
\end{equation}

\section{The two-point correlator}
The two-point correlator can be constructed as
\begin{equation}
\Pi_{ij}(\omega=p\cdot v)=i\int\langle 0|T\{J_i(x)\bar J_j(0)\}|0\rangle
  e^{ipx}d^4x
  =\Gamma'\frac{1+\slv}2\bar\Gamma'\frac14{\rm tr}(\Gamma\bar\Gamma)
  2{\rm tr}(\tau\tau^\dagger)P_{ij}(\omega).
\end{equation}
$i$ and $j$ stand for the two independent current representations for the
ground state baryons. This gives rise to two independent types of
correlators, namely the diagonal correlators $\langle J_1\bar J_1\rangle$
and $\langle J_2\bar J_2\rangle$ and the non-diagonal correlators
$\langle J_1\bar J_2\rangle$ and $\langle J_2\bar J_1\rangle$ as well as
linear combinations of these cases. Following the standard QCD sum rule
method~\cite{rostock6}, the correlator is calculated in the Euclidean 
region $-\omega\approx 1-2\GeV$ including perturbative and non-perturbative 
contributions. In the Euclidean region the non-perturbative contributions 
are expected to form a convergent series. The non-perturbative effects are 
taken into account by employing an operator product expansion (OPE) for the 
time-ordered product of the currents. One then has
\begin{eqnarray}
\langle T\{J(x),\bar J(0)\}\rangle&=&\sum_dC_d(x^2)O_d\ =\\
  &=&C_0(x^2)O_0+C_3(x^2)O_3+C_4(x^2)O_4+C_5(x^2)O_5+\ldots\nonumber
\end{eqnarray}
where the $O_d$ are vacuum expectation values of local operators whose
mass dimensions are labelled by their subscripts $d$. $O_0=\hat 1$
corrersponds to the so-called perturbative term,
$O_3=\langle\bar qq\rangle$ is a quark condensate term,
$O_4=\as\langle G^2\rangle$ is a gluon condensate term,
$O_5=g_s\langle\bar q\sigma_{\mu\nu}G^{\mu\nu}q\rangle$ is a mixed
quark-gluon condensate, and so on. The expansion coefficient $C_d(x^2)$ are
the corresponding Wilson coefficients of the operator product expansion. The
different diagrams are shown in Fig.~1.
\medskip\\
A straightforward dimensional analysis shows that the operator product
expansion of the diagonal correlators contain only even-dimensional terms,
while the expansion of the non-diagonal correlators contain only
odd-dimensional terms. This classification is preserved when radiative
corrections are included, assuming the light quarks to be massless. 

\section{Construction of QCD sum rules}
The QCD sum rules can be constructed by taking care of two possible
expressions for the two-point correlator. On the one hand, the scalar
correlator function $P(\omega)$ satisfies the dispersion relation
\begin{equation}
P(\omega)=\int_0^\infty\frac{\rho(\omega')d\omega'}{\omega'-\omega-i0}
  +\mbox{subtraction},
\end{equation}
where $\rho(\omega)={\sl Im}(P(\omega))/\pi$ is the spectral density. On the
the phenomenological side of view, the two-point correlator is represented
by the spectral representation
\begin{equation}
\Pi(\omega)=\frac12\sum_X\frac{|\langle 0|J_X|X\rangle|^2}{\omega_X-\omega
  -i0}+\hbox{subtraction}.
\end{equation}
The scalar correlator funktion can thus be expressed by the {\em residues
$F_B$} given by
\begin{equation}
\langle 0|J_\Lambda|\Lambda_Q\rangle=F_\Lambda u,\quad
\langle 0|J_\Sigma|\Sigma_Q\rangle=F_\Sigma u\quad\mbox{and}\quad
\langle 0|J_{\Sigma^*}^\nu|\Sigma_Q^*\rangle=\frac1{\sqrt3}F_{\Sigma^*}u^\nu
\end{equation}
in the form
\begin{equation}
P(\omega)=\frac{\frac12|F_B|^2}{\bar\Lambda-\omega-i0}
  +\sum_{X\ne B}\frac{\frac12|F_X|^2}{\omega_X-\omega-i0}
  +\mbox{subtraction},
\end{equation}
where $\bar\Lambda=m_B-M_Q$ is the ground state energy of the baryon. The
main assumption of the QCD sum rule method is that the remaining sum can be
approximated by the integral of the spectral density given by the dispersion
relation starting from some threshold energy $E_C$. The combination of the
phenomenological and the theoretical identity for the correlator function
leads to
\begin{equation}
\frac{\frac12|F_B|^2}{\bar\Lambda-\omega-i0}=\int_0^{E_C}
  \frac{\rho(\omega')d\omega'}{\omega'-\omega-i0}+\mbox{subtraction}.
\end{equation}
This formula is not useful anyhow, since the spectral density is reliable
only for negative values of $\omega$, while the integral is to be calculated
mainly at $\omega=\bar\Lambda$. This region of integration can be reached by
an extrapolation using higher and higher derivatives when $\omega$ goes to
$-\infty$. This extrapolation is expressed by the {\em Borel transformation}
\begin{equation}
\hat Bf(T)=\hat B_T^{(\omega)}(f(\omega))
  :=\lim_{\-\omega,n\rightarrow\infty}\frac{(-\omega)^{n+1}}{n!}
  \frac{d^n}{d\omega^n}f(\omega),\qquad T=\frac{-\omega}n\quad\mbox{fixed}.
\end{equation}
This Borel transformation leads to the final form of the QCD sum rules,
namely
\begin{equation}
\frac12|F_B|^2e^{-\bar\Lambda/T}=\int_0^{E_C}\rho(\omega)e^{-\omega/T}
  d\omega.
\end{equation}
The Borel transformation also cancels the subtraction term. The constant
ratio $T$ is called {\em Borel parameter}. It is an unphysical quantity in
units of an energy, and the obtained values should be mostly independent on
this parameter. This is the main criterion in analyzing the sum rules.

\section{Diagonal and non-diagonal correlators}
I don't want to go into details concerning the calculations of first order
radiative corrections to the two-point correlators (for details 
see~\cite{rostock2,rostock3}). We calculate radiative corrections only to 
the leading terms in the operator product expansion because the non-leading 
contributions are small. It should be stressed that we could make use of 
the same MATHEMATICA package which was constructed for the calculation of 
two-loop corrections of the currents itself. For the case of the diagonal 
correlator (for an overview over the contributing diagrams see Fig.~2) we 
ended up with a renormalized spectral density
\begin{equation}
\rho_0^{\rm ren}(\omega)=\rho_0^{\rm Born}(\omega)
  \left[1+\frac\as{4\pi}r(\omega/\mu)\right],
\end{equation}
where
\begin{equation}
\rho_0^{\rm Born}(\omega)=\frac{\omega^5}{20\pi^4}\quad\mbox{and}\quad
  r(\omega/\mu)=r_1\ln\left(\frac\mu{2\omega}\right)+r_2
\end{equation}
with
\begin{equation}
r_1=\frac83(n_\gamma^2-4n_\gamma+6)\quad\mbox{and}\quad
r_2=\frac8{45}(38n_\gamma^2-137n_\gamma+273+60\zeta(2)),
\end{equation}
which results in the sum rule
\begin{eqnarray}
\frac12F^2(\mu)e^{-\bar\Lambda/T}&=&\frac{N_c!}{\pi^4}
  \bigg[T^6\left(f_5(x_C)+\frac\as{4\pi}\left(\left(\ln\frac\mu{2T}\right)
  f_5(x_C)-f_5^l(x_C)\right)r_1+f_5(x_C)r_2\right)\,+\nonumber\\
  &&+cE_G^4T^2f_1(x_C)+E_Q^6\exp\left(-\frac{2E_0^2}{T^2}\right)\bigg],
\end{eqnarray}
where $c$ is a Clebsch-Gordan type factor which takes the values $c=1$ for
the $\Lambda_Q$-type and $c=-1/3$ for the $\Sigma_Q$-type doublet
$\{\Sigma_Q,\Sigma_Q^*\}$ ground state baryons. In order to simplify the
notation I have introduced the abbreviations
\begin{equation}
x_C:=\frac{E_C}T,\quad E_0:=\frac{m_0}4,\quad
  (E_Q)^3:=-\frac{\pi^2}{2N_C}\langle\bar qq\rangle\quad\mbox{and}\quad
  (E_G)^4:=\frac{\pi\as\langle G^2\rangle}{32N_c(N_c-1)}.
\end{equation}
The sum rule is expressed in terms of the modified Gamma functions
\begin{equation}
f_n(x):=\int_0^x\frac{x'^n}{n!}e^{-x'}dx'
  =1-e^{-x}\sum_{m=0}^n\frac{x^m}{m!},\qquad
f_n^l(x):=\int_0^x\frac{x'^n}{n!}\ln x'e^{-x'}dx'.
\end{equation}
I have omitted the index for the current. Later on we will see that the
sum rule analysis is indeed independent of the chosen current within the
assumed error bars. So we can use only one residue also for the non-diagonal
case. For the non-diagonal correlator (cf.\ Fig.~3) we obtain a spectral
density part
\begin{equation}
\rho_3^{\rm ren}(\omega)=\rho_3^{\rm Born}(\omega)
  \left[1+\frac\as{4\pi}r(\omega/\mu)\right],
\end{equation}
where
\begin{equation}
\rho_3^{\rm Born}(\omega)=-\frac{\langle\bar qq\rangle^{\rm ren}}{\pi^2}
  \omega^2\quad\mbox{and}\quad
  r(\omega/\mu)=r_1\ln\left(\frac\mu{2\omega}\right)+r_2
\end{equation}
with
\begin{eqnarray}
r_1&=&\frac43(2n_\gamma^2-8n_\gamma+7+2(n_\gamma-2)s_\gamma)\quad\mbox{and}
  \nonumber\\
r_2&=&\frac23(8n_\gamma^2-28n_\gamma+37+8n_\gamma s_\gamma-12s_\gamma
  +8\zeta(2)).
\end{eqnarray}
Here the sum rule has the form
\begin{eqnarray}
\frac12F^2(\mu)e^{-\bar\Lambda/T}\!&=&\!\frac{2N_c!}{\pi^4}
  \bigg[E_Q^3T^3\left(f_2(x_C)+\frac\as{4\pi}
  \left(\left(\ln\left(\frac\mu{2T}\right)f_2(x_C)-f_2^l(x_C)\right)r_1
  +f_2(x_C)r_2\right)\right)\nonumber\\&&
  -E_Q^3E_0^3T\left(1-\frac c2\right)f_0(x_C)
  +\frac23\left(1-\frac c2\right)\frac{E_Q^3E_G^4}T+\frac{\as C_F}{36\pi}
  \frac{E_Q^9}{T^3}\bigg].
\end{eqnarray}

\section{Numerical analysis}
Having the neccessary formulae at hand we can start the numerical analysis 
of the sum rules. In doing this we use the following numerical input values 
for the condensate contributions~\cite{rostock6,rostock7}
\begin{eqnarray}\label{condensates}
\langle\bar qq\rangle&=&-(0.23\GeV)^3\quad \mbox{(quark condensate)},\\
\as\langle G^2\rangle&=&0.04\GeV^4 \quad \mbox{(gluon condensate)},\quad
  \mbox{and}\nonumber\\
g_s\langle\bar q\sigma_{\mu\nu}G^{\mu\nu}q\rangle&=&m_0^2\langle\bar qq\rangle
  \quad\mbox{with}\quad m^2_0=0.8\GeV^2\quad
  \mbox{(mixed quark-gluon condensate)}.\nonumber
\end{eqnarray}
There are in general two strategies for the numerical analysis of the QCD 
sum rules. The first strategy fixes the bound state energy $\bar\Lambda$ 
from the outset by choosing a specific value for the pole mass of the heavy 
quark and then extracts a value for the residue $F$. In order to obtain 
information from the sum rules which is independent of specific input 
values, we adopt a second strategy, namely to determine both $\bar\Lambda$ 
and $F$ by finding simultaneous stability values for them with respect to 
the Borel parameter $T$.

The first step in carrying out the numerical analysis of the sum rules 
is to find a sum rule ``window'' for the allowed values of the Borel 
parameter $T$. The parameter range of $T$ is constrained by two different 
physical requirements. The first is that the convergence of the OPE 
expansion must be secured. We therefore demand that the subleading term 
in the OPE does not contribute more than $30\%$ of the leading order term. 
This gives a lower limit for the Borel parameter. The upper limit is 
determined by the requirement that the contributions from the excited 
states plus the physical continuum (even after Borel transformation) 
should not exceed the bound state contribution. This requirement is 
neccessary in order to guarantee that the sum rules are as independent as 
possible of the model-dependent assumptions concerning the profile of the 
theoretical spectral density, i.e.\ the model of the continuum. The lower 
limit of $E_C$ is given by the requirement that the indicated window should 
be kept open. For the rest, $E_C$ is a free floating variable which is only 
limited by the stability requirements on $\bar\Lambda$ and $F$.
\medskip\\
As an example I want to show the analysis of the diagonal sum rule for the 
$\Lambda_Q$-type state in Fig.~4. Fig~4(a) shows the dependence of the bound 
state energy $\bar\Lambda$ on the Borel parameter $T$ and Fig.~4(b) shows 
the dependence of the residue on $T$, both for the leading order sum rule. 
Fig.~4(c) and Fig.~4(d) show the same dependencies for the radiatively 
corrected sum rules. The same analysis is repeated for the $\Sigma_Q$-type 
baryons. The results of the numerical analysis both without and with 
radiative corrections are given in Table~\ref{tab1}.
\begin{table}\begin{center}
\begin{tabular}{|c||c|c|c|}
Baryon type state&$E_C$&$\bar\Lambda$&$F$\\\hline
$\Lambda_Q$ (L.O.)
  &$1.2\pm 0.1\GeV$&$0.77\pm 0.05\GeV$&$0.022\pm 0.001\GeV^3$\\
$\Lambda_Q$ (N.L.O.)
  &$1.1\pm 0.1\GeV$&$0.77\pm 0.05\GeV$&$0.027\pm 0.002\GeV^3$\\
$\Sigma_Q$ (L.O.)
  &$1.4\pm 0.1\GeV$&$0.96\pm 0.05\GeV$&$0.031\pm 0.002\GeV^3$\\
$\Sigma_Q$ (N.L.O.)
  &$1.3\pm 0.1\GeV$&$0.94\pm 0.05\GeV$&$0.038\pm 0.003\GeV^3$\\
\hline\end{tabular}
\caption{\label{tab1}Results of the diagonal sum rule analysis for the 
continuum threshold parameter $E_C$, the bound state energy $\bar\Lambda$, 
and the residuum $F$ for $\Lambda_Q$-type and $\Sigma_Q$-type currents, 
analyzed to leading order (L.O.) as well as next-to-leading order (N.L.O.)}
\end{center}\end{table}
\medskip\\
The numerical results for the non-diagonal sum rules are given in 
Table~\ref{tab2}. Assuming relative errors of $10\%$ for the bound state 
energy and $20\%$ for the residue, the obtained values are in agreement with 
the results of the analysis of the diagonal sum rules, where the values for 
the $\Sigma_Q$-type baryon are the more reliable one.
\begin{table}[ht]\begin{center}
\begin{tabular}{|c||c|c|c|}
Baryon type state&$E_C$&$\bar\Lambda$&$F$\\\hline
$\Lambda_Q$ (L.O.)
  &$1.0\pm 0.10\GeV$&$0.75\pm 0.10\GeV$&$0.024\pm 0.002\GeV^3$\\
$\Lambda_Q$ (N.L.O.)
  &$1.0\pm 0.10\GeV$&$0.72\pm 0.10\GeV$&$0.032\pm 0.003\GeV^3$\\
$\Sigma_Q$ (L.O.)
  &$1.5\pm 0.10\GeV$&$1.16\pm 0.10\GeV$&$0.045\pm 0.003\GeV^3$\\
$\Sigma_Q$ (N.L.O.)
  &$1.2\pm 0.10\GeV$&$0.94\pm 0.10\GeV$&$0.039\pm 0.004\GeV^3$\\
\hline\end{tabular}
\caption{\label{tab2}Results of the non-diagonal sum rule analysis for the 
continuum threshold parameter $E_C$, the bound state energy $\bar\Lambda$, 
and the residuum $F$ for $\Lambda_Q$-type and $\Sigma_Q$-type currents, 
analyzed to leading order (L.O.) as well as next-to-leading order (N.L.O.)}
\end{center}\end{table}
\begin{table}[ht]\begin{center}
\begin{tabular}{|c||c|c|c|}
Baryon type state&$E_C$&$\bar\Lambda$&$F$\\\hline
$\Lambda_Q$ (L.O.)
  &$1.1\pm 0.10\GeV$&$0.77\pm 0.10\GeV$&$0.034\pm 0.004\GeV^3$\\
$\Lambda_Q$ (N.L.O.)
  &$1.1\pm 0.10\GeV$&$0.77\pm 0.10\GeV$&$0.032\pm 0.004\GeV^3$\\
$\Sigma_Q$ (L.O.)
  &$1.3\pm 0.10\GeV$&$1.03\pm 0.10\GeV$&$0.045\pm 0.004\GeV^3$\\
$\Sigma_Q$ (N.L.O.)
  &$1.2\pm 0.10\GeV$&$0.94\pm 0.10\GeV$&$0.036\pm 0.004\GeV^3$\\
\hline\end{tabular}
\caption{\label{tab3}Results of the constituent type mixed sum rule analysis 
for the continuum threshold parameter $E_C$, the bound state energy 
$\bar\Lambda$, and the residuum $F$ for $\Lambda_Q$-type and $\Sigma_Q$-type 
currents, analyzed to leading order (L.O.) as well as next-to-leading order 
(N.L.O.)}
\end{center}\end{table}
\medskip\\
For the linear combination $J=(J_1+J_2)/2$ of currents the light-side Dirac 
structure can be seen to appear in the form $\frac12(1+\slv)\Gamma$, i.e.\ 
one has the projector factor $P_+=(1+\slv)/2$ which projects on the large 
components of the light quark fields. We refer to this particular linear 
combination of currents as the {\em constituent type current}. This linear 
combination of currents is expected to have minimum overlap with the heavy 
ground state baryons in the constituent quark model, i.e.\ where the light 
diquark state in the heavy baryon is taken to be composed of on-shell light 
quarks. This model emerges in the large $N_c$-limit.
\medskip\\
The use of a constituent type interpolating current $J=(J_1+J_2)/2$ 
combines the two sum rule formulas for the diagonal and the non-diagonal 
case, taking one half of each part. The numerical results of the analysis 
are given in Table~\ref{tab3}.
The constituent type sum rules show an improved stability on the Borel 
parameter $T$ as compared to the non-diagonal sum rules, but the stability 
is not as good as in the diagonal case. Within the assumed errors the 
results are again in agreement with both the diagonal and the non-diagonal 
sum rule analysis.

\section{Comparison with experimental values}
Taking the experimental results for charm-quark baryons, namely 
$m(\Lambda_c)=2284.9\pm 0.6\MeV$ and $m(\Sigma^+_c)=2453.5\pm 0.9\MeV$, our 
central values $\bar\Lambda(\Lambda_Q)=760\MeV$ and 
$\bar\Lambda(\Sigma_Q)=940\MeV$ predict a mean pole mass of $m_c=1520\MeV$ 
for the charm quark. For the experimental value $m(\Lambda_b)=5642\pm 50\MeV$ 
for the mass of the baryon $\Lambda_b$, our central value 
$\bar\Lambda(\Lambda_Q)=760\MeV$ for the bound state energy suggests a pole 
mass of $m_b=4880\MeV$ for the bottom quark.

\newpage

\section{Conclusion and outlook}
\begin{itemize}
\item The determination of the two-loop anomalous dimension is essential 
for the evolution behaviour of the current to first order perturbation 
theory.
\item The presented results as well as the related computer package are 
applicable and extendable to the calculation of three-point correlators and 
thus the determination of the Isgur-Wise function. This work will be 
started in the near future.
\item Another extend is the calculation of correlators including massive 
lines within exact QCD. This will be done to test the convergence of the 
$1/m_Q$ expansion in HQET.
\end{itemize}

\newpage

\vspace{1cm}
\centerline{\Large\bf Figure Captions}
\vspace{.5cm}
\newcounter{fig}
\begin{list}{\bf\rm Fig.\ \arabic{fig}:}{\usecounter{fig}
\labelwidth1.6cm\leftmargin2.5cm\labelsep.4cm\itemsep0ex plus.2ex}

\item Diagrams representing the first few vacuum expectation values of local 
operators due to the operator product expansion of the two-point correlator.

\item Radiative corrections to the diagonal correlator.\\
(0) Lowest order two-loop contribution,\\
(1)--(4) $O(\as)$ three-loop contributions.

\item Radiative corrections to the non-diagonal correlator given by the 
dimension three condensate contribution.\\
(0) Lowest order one-loop contribution,\\
(1)--(4) $O(\as)$ two-loop contributions.

\item Sum rule results on the non-perturbative parameters of the $\Lambda_Q$ 
as functions of the Borel parameter $T$. Shown are five curves for five 
different values of the threshold energy $E_C$ spaced by $100\MeV$ around 
the central value $E_C=E_C^{\rm best}$. $E_C$ grows from bottom to top. 
These are in detail\\
(a) lowest order sum rule results for the bound state energy 
$\bar\Lambda(\Lambda)$;\\
(b) lowest order sum rule results for the absolute value of the residue 
$F_\Lambda$;\\
(c) $O(\as)$ sum rule results for the bound state energy 
$\bar\Lambda(\Lambda)$;\\
(d) $O(\as)$ sum rule results for the absolute value of the residue 
$F_\Lambda$.

\end{list}

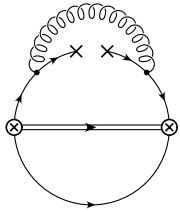
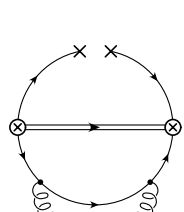
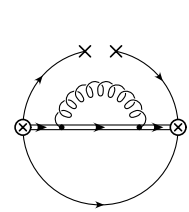
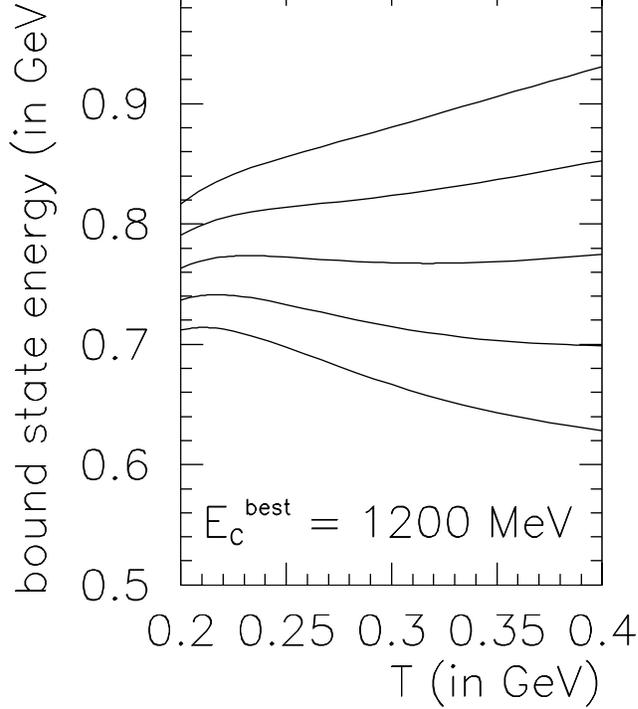
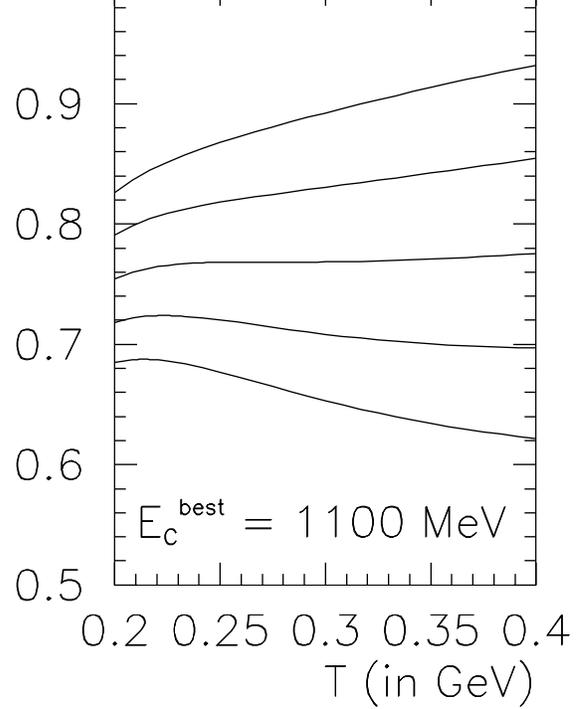
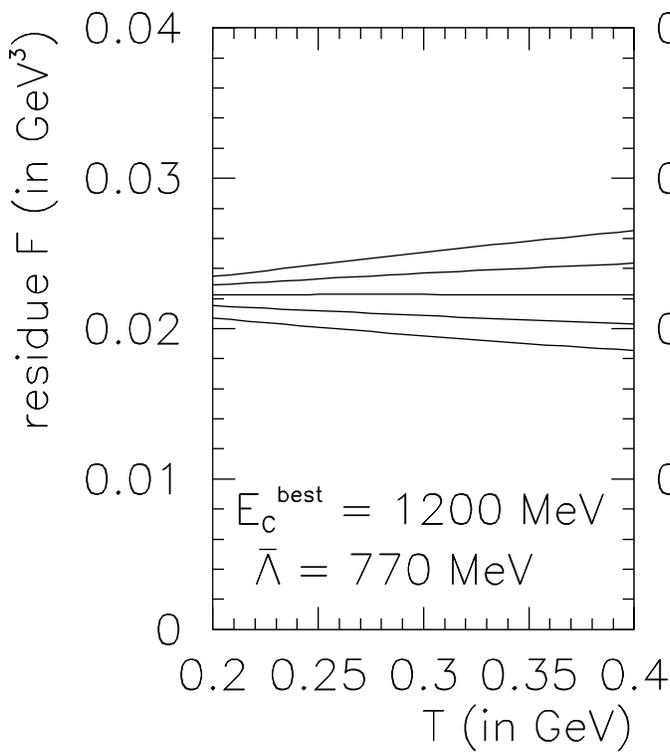
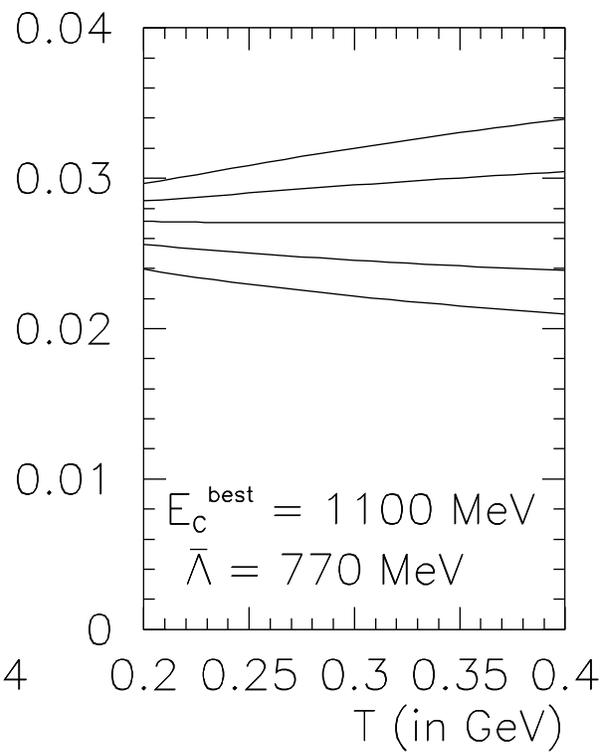
\end{document}